\documentclass[twocolumn,prb,floatfix,showpacs]{revtex4}

\usepackage{graphicx}
\usepackage{epstopdf} 
\usepackage{dcolumn}
\usepackage{bm}
\usepackage{hyperref}

\usepackage{siunitx} 

\usepackage{braket}
\usepackage{amsmath}
\usepackage{multirow}
\usepackage{booktabs}
\usepackage{color}
\usepackage{ulem} 
\usepackage{scrextend}

\usepackage[utf8]{inputenc}

\begin{document}


\title{Competing Ferromagnetic and Anti-Ferromagnetic interactions in Iron Nitride $\zeta$-Fe$_2$N} 



\author{K. Sandeep Rao}
\author{H. G. Salunke}%
 \email{hsalunke@barc.gov.in}
\affiliation{%
 Technical Physics Division, Bhabha Atomic Research Centre, Mumbai.
}%


\date{\today}

\begin{abstract}
The paper discusses the magnetic state of zeta phase of iron nitride viz. $\zeta$-Fe$_2$N on the basis of spin polarized first principles electronic structure calculations together with a review of already published data. Results of our first principles study suggest that the ground state of $\zeta$-Fe$_2$N is ferromagnetic (FM) with a magnetic moment of 1.528 $\mu_\text{B}$ on the Fe site. The FM ground state is lower than the anti-ferromagnetic (AFM) state by 8.44 meV and non-magnetic(NM) state by 191 meV per formula unit. These results are important in view of reports which claim that $\zeta$-Fe$_2$N undergoes an AFM transition below 10K and others which do not observe any magnetic transition up to 4.2K. We argue that the experimental results of AFM transition below 10K are inconclusive and we propose the presence of competing FM and AFM superexchange interactions between Fe sites mediated by nitrogen atoms, which are consistent with Goodenough-Kanamori-Anderson rules. We find that the anti-ferromagnetically coupled Fe sites are outnumbered by ferromagnetically coupled Fe sites leading to a stable FM ground state. A Stoner analysis of the results also supports our claim of a FM ground state. 
\end{abstract}


\maketitle

\section{Introduction}
Nitriding is a process used for surface hardening of austenitic steel which improves the corrosion and wear resistance of its surface.\cite{nitrocarb} It involves passing a gaseous mixture of ammonia and hydrogen on Fe surface, which produces a number of iron compounds, of the type Fe$_x$N, depending upon the temperature and nitrogen chemical potential at the surface of iron.  Most of these iron nitride phases are unstable at ambient temperatures and pressures but the high kinetic barrier of the $N + N \rightarrow N_2$ reaction prevents their decomposition at temperatures below 400 \si{\degree}C. \cite{lehrer_2002} Hence, nitrified steel at ambient temperatures and pressures could contain phases of the type Fe$_4$N, Fe$_2$N etc. on its surface which would greatly affect its properties and concomitant applicability. Very recently, Fe$_{16}$N$_2$ was in news as strongest known magnet yet and as a candidate for magnetic motors.\cite{jiang_synthesis_2016} This together with search for materials with potential magnetic properties has brought iron nitrides to the attention of materials research community. 

Magnetic state of Fe$_2$N, both experimentally and theoretically, has been a topic of debate since the beginning of its investigation. The experimental measurements by Bridelle \cite{bridelle} suggested feeble FM magnetization in $\zeta$-Fe$_2$N at 78K, while a later report by Bainbridge et al.\cite{bainbridge_mossbauer_1973} based on M{\"o}ssbauer spectra did not observe any magnetic splitting in the material down to 4.2K. Thus, they reported a paramagnetic(PM) state of Fe atoms in the material. Subsequent experimental works on the material have only added to the controversy, as findings by Hinomura et. al \cite{hinomura_study_1998} and Kano et. al. \cite{kano_magnetism_2001} have suggested that the material is AFM with a N{\' e}el temperature in the range of about 9K-11K. One of the major hurdles to experimental investigation of $\zeta$ phase can be traced back to its strict stoichiometric composition of Fe$_2$N and its structural similarity to $\epsilon$ phase which is non-stoichiometric with a composition Fe$_{2-3}$N.

Theoretical first principles calculations have also been inconsistent regarding the magnetic state of $\zeta$-Fe$_2$N, with different publications reporting conflicting results. Matar and Mohn\cite{matar_electronic_1993} report a FM state that is more stable than an unpolarized state by 175 meV per formula unit and magnetic moment per Fe atom of 1.47 $\mu_\text{B}$. Their Density of States (DOS) curves show a metallic state and itinerant ferromagnetism. Additionally, an analysis on the basis of Slater-Pauling-Friedel plot predicts that $\zeta$-Fe$_2$N should behave like a strong ferromagnet. The work of Eck and coworkers \cite{eck_theoretical_1999}, shows a metallic phase with 1.5 unpaired electrons per Fe atom. They report that the phase is PM. Further work of Sifkovits et. al. \cite{sifkovits_interplay_1999}, reports a magnetic moment of 1.43$\mu_\text{B}$ per iron atom. The DOS reported shows a metallic state with Fermi Level(E$_F$) lying in Fe band. They do not claim PM or FM state for $\zeta$-Fe$_2$N. Thereafter, comprehensive investigations of electronic structure of several phases of Fe$_2$N and Fe$_2$C were studied in a recently published work by Fang and co-workers\cite{fang_role_2011}. The work dealt with understanding relative stability, bonding character and Fermi level properties of several possible structures of Fe$_2$N and Fe$_2$C. Out of the several Fe$_2$N structures possible, relative enthalpy of formation is observed to be highest for $\zeta$-Fe$_2$N structure and is likely to be the most probable structural phase formed on the surface of nitrided steel. Fang's works yield a magnetic moment of 1.56 $\mu_\text{B}$ per iron atom. They claim that the structure is FM. More recently, calculations by Chen et al\cite{chen_phase_2015}, yield 1.38 unpaired electrons per Fe. They claim that the structure is PM and the DOS curve shows a metallic state. The results of Chen are intriguing considering they obtain a magnetic moment for Fe based on spin polarized calculation and that estimates of Stoner parameter are not reported by them.

Considering immense potential magnetic applications for iron nitrides and the lack of a clear understanding, both experimentally and theoretically, with regards to their magnetic state, we undertake detailed investigations into the magnetic state of $\zeta$-Fe$_2$N structure. The next section describes the crystal structure and methodology employed while section III discusses results of electronic structure calculations. In section IV we discuss magnetism in this structure and section V concludes the work.

\section{Methodology}
\begin{figure}
\includegraphics[width=8.6cm]{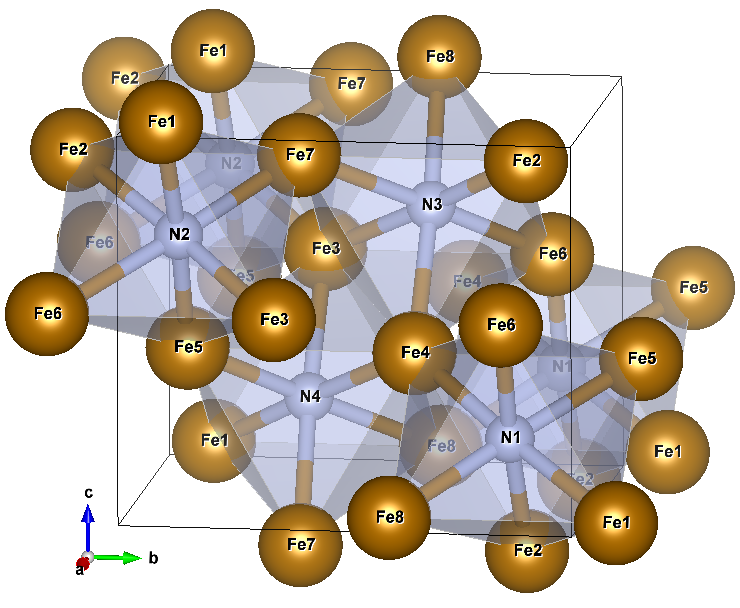}
\caption{(Color online) Orthorhombic unit cell of $\zeta$-Fe$_2$N.}
\label{fig:unitcell}
\end{figure}

$\zeta$-Fe$_2$N has an orthorhombic unit cell, with space group Pbcn(\#60), where the Fe and N atoms occupy 8d and 4c crystallographic positions respectively (see Table \ref{table:chargesmom}). The structure may be considered as a distorted HCP lattice of Fe atoms with N atoms occupying octahedral sites. A nitrogen atom is bonded to 6 Fe atoms, with 2 atoms each at a distance of 1.87 \AA, 1.92 \AA \ and 1.96 \AA, giving rise to slightly tetragonally distorted octahedra. 
In view of discussion on magnetic state in later sections, we note from FIG. \ref{fig:unitcell}, a Fe atom is at the vertex of three tetrahedra and is hence bonded to three N-atoms and has fifteen nearest neighbor Fe atoms. Out of these fifteen Fe-Fe nearest neighbor pairs, three pairs have Fe-N-Fe bond angle of $\sim$ 180 $^o$ while the remaining 12 pairs have a Fe-N-Fe bond angle of $\sim$ 90 $^o$.\footnote{The $\sim$ 180 $^o$ angles are actually, 177.15, 177.14 and 177.15 degrees while the $\sim$ 90 $^o$ ones are 87.8, 90, 86.1, 91, 87.8, 92.02, 89.9, 90.1, 91.06, 91.77, 89.96, 92.04 degrees.} For example in FIG. \ref{fig:unitcell}, consider the atom Fe3: It has three $180 ^o$ neighbors (namely Fe3-N3-Fe2, Fe3-N4-Fe7, Fe3-N2-Fe2) and twelve $90 ^o$ neighbors (i.e. the rest of the Fe atoms in the three octahedra).

All electronic structure calculations were performed by using the density functional theory codes implemented in Vienna ab-initio simulation package (VASP)\cite{kresse_ab_1993}. The Projector Augmented Wave (PAW) method was employed and exchange correlation interactions were treated by means of Perdew-Burke-Ernzerhof (PBE) functional under the generalized gradient approximation (GGA). The PAW potentials were constructed for valence configurations: Fe(4s$^1$3d$^7$) and N(2s$^2$2p$^3$). The electron occupations were smeared with a gaussian distribution function with a smearing width of 1 meV. The geometry optimization were performed with a converged Monkhorst Pack grid of (9x9x9) k-points till the residual forces were less than 0.01 eV\AA$^{-1}$ and tolerances of 10$^{-5}$ eV was chosen for self-consistency loops. The plane waves were incorporated till a high energy cut-off of 400 eV. The DOS calculations were performed using tetrahedron method with Bl{\"o}chl corrections.

For benchmarking the results of Fe$_2$N, first principles calculations have also been performed on  for bcc-Fe using a dense mesh of 21x21x21 k-points with other convergence and tolerance values kept similar to the Fe$_2$N ones. The converged lattice constant and magnetic moment per atom were 2.832 \AA \  and 2.186$\mu_\text{B}$, which are in excellent agreement with previously reported\cite{kubler_magnetic_1981} values of 2.82\AA \  and 2.18 $\mu_\text{B}$.

Crystal Overlap Hamilton Population (COHP) analysis was done using the LOBSTER\cite{dronskowski_crystal_1993} package. Bader charge analysis was done using the algorithm implemented by Henkelman\cite{tang_grid-based_2009} research group. All crystal structure and charge density images were rendered using VESTA\cite{momma_vesta_2008} software package.

\section{Results}
Unit cell volume relaxations were done and the converged lattice parameters (a=4.3466 \AA, b=5.4682 \AA, c=4.7436 \AA) obtained via spin polarised calculations are close to previously reported experimental values by Rechenbach and Jacobs\cite{rechenbach_structure_1996} (a=4.437 \AA, b=5.541 \AA, c=4.843 \AA) and computational work done by Fang\cite{fang_role_2011} et al (a=4.3406 \AA, b=5.4480 \AA, c=4.7544 \AA).

Total energy calculations show that the FM ground state is more stable than the NM state and AFM state by 191 meV and 8.44 meV per formula unit respectively. All results reported henceforth pertain to the FM state.

The charges on ions were computed using Bader charge analysis and are tabulated in Table \ref{table:chargesmom} which clearly shows a charge transfer of 0.8e$^{-}$ occurring from Fe to N. This implies that 2 Fe lose their 4s electrons to 2p state of N. This charge transfer should cause the Fe 4s band to lie in the conduction band with the N 2p band partially filled and hybridizing with the 3d band of Fe.

Our charge transfer results are in agreement with Fang et al's report of 0.62 e$^-$/Fe atom being transferred to the more electronegative N atom but are contrary to the nitrogen donor model proposed by Bainbridge et al \cite{bainbridge_mossbauer_1973}. According to it, 3 N atoms surrounding each Fe donate 1.5 electrons and thus predicts a Fe electronic configuration 3d$^{8.6}$ 4s$^{0.9}$ with an expected magnetic moment of 1.5 $\mu_\text{B}$. Matar and Mohn \cite{matar_electronic_1993} also report a small amount of charge transfer occurring from N to Fe but report the bonding to be primarily covalent in nature.

\begin{table}
\caption{Charges and magnetic moments on ions obtained by Bader analysis.}
\label{table:chargesmom}
\begin{tabular}{|c|c|c|c|}
\hline 
\multirow{2}{*}{Atom} & \multirow{2}{*}{Wyckoff Site} & \multirow{2}{*}{Charge (e)} & Magnetic \tabularnewline
 &  &  & moment ($\mu_B$)\tabularnewline
\hline 
\hline 
Fe & 8d (0.251, 0.128, 0.008) & +0.834 & 1.528\tabularnewline
\hline 
N & 4c (0, 0.864, 0.25) & -1.668 & -0.06\tabularnewline
\hline 
\end{tabular}
\end{table}

\begin{figure*}
\includegraphics[width=17.2cm]{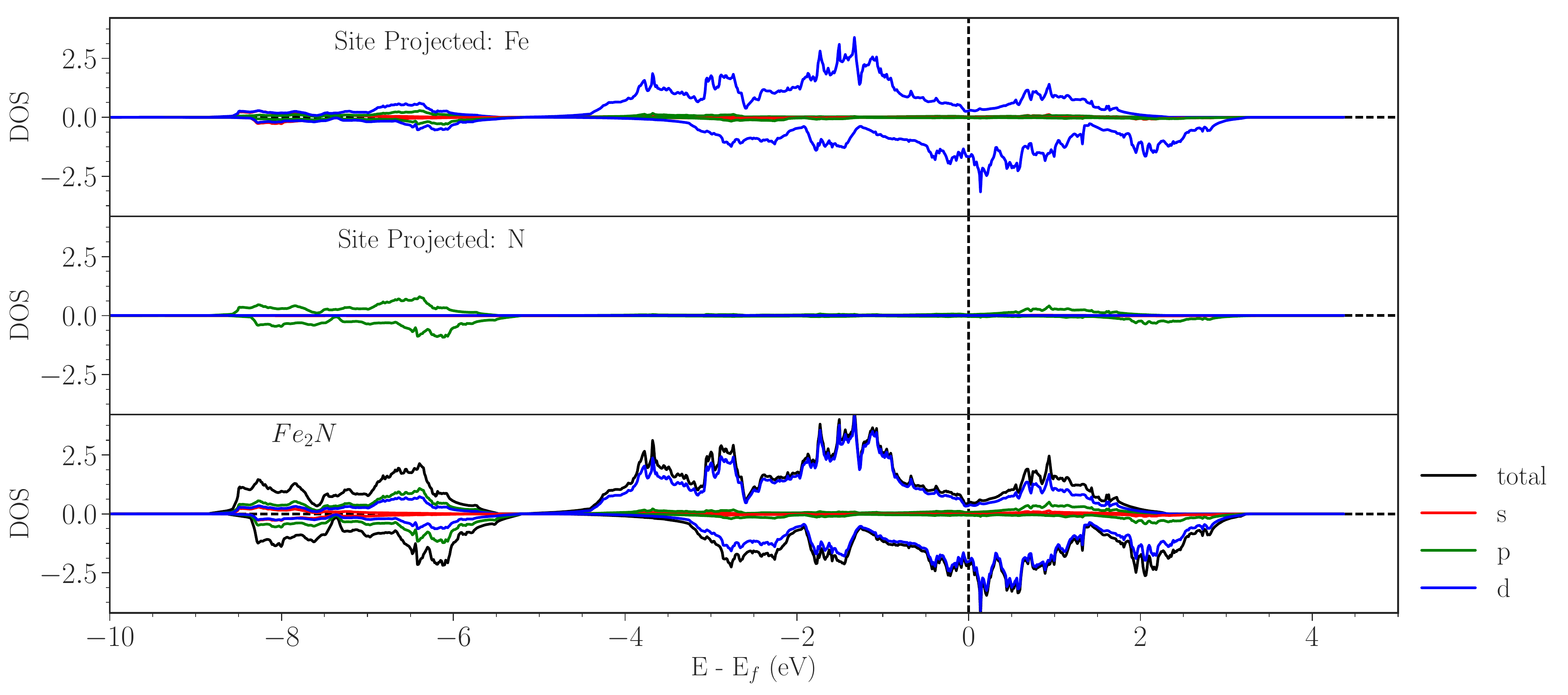}
\caption{(Color online) Site and orbital projected spin polarized DOS of $\zeta$-Fe$_2$N. DOS in states/eV/formula units}
\label{fig:DOS}
\end{figure*}

\begin{figure*}
\includegraphics[width=18cm]{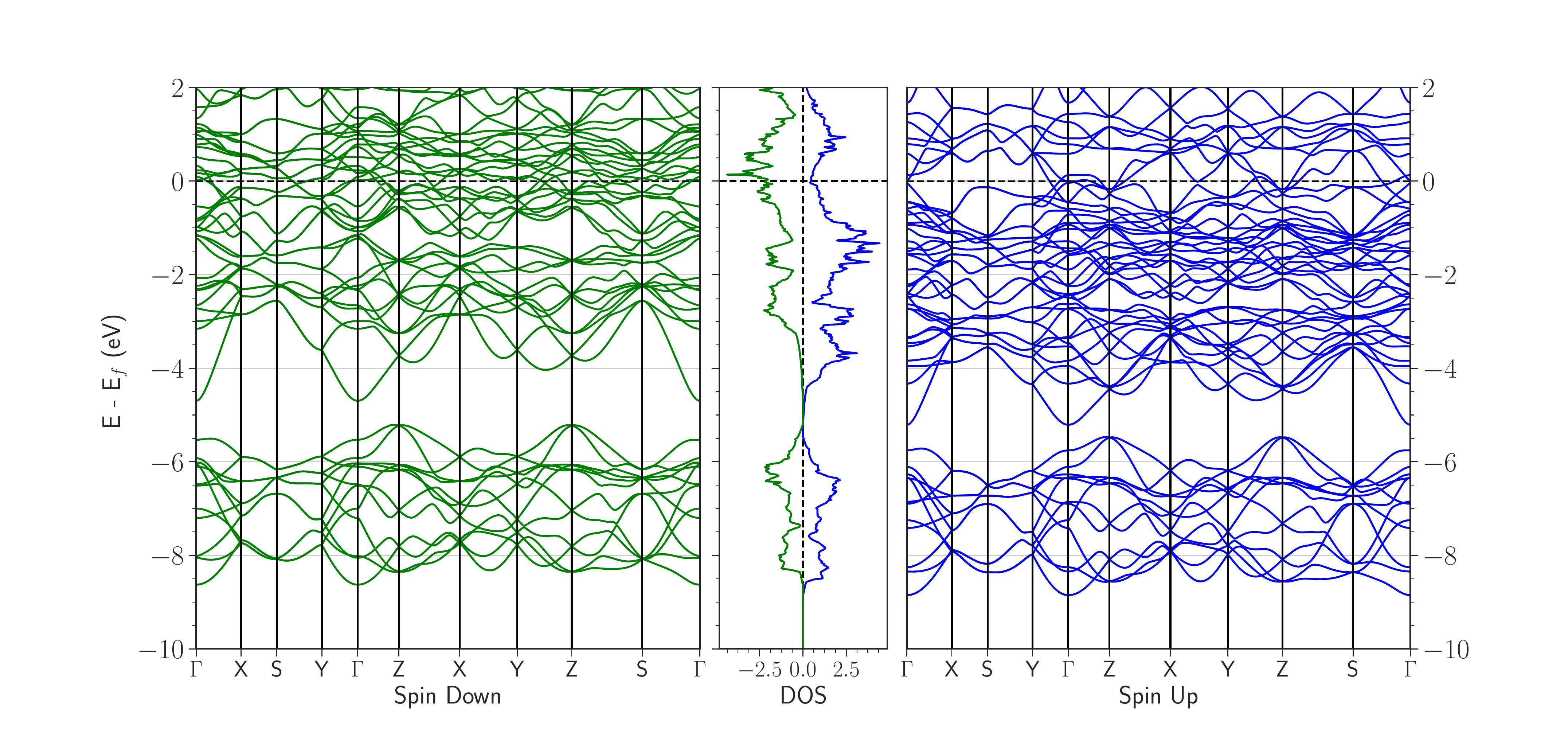}
\caption{(Color online) Spin polarized band structure and total density of states for unit cell of $\zeta$-Fe$_2$N}
\label{fig:bands}
\end{figure*}

These effects of charge transfer are apparent in the site and orbital projected DOS curve depicted in FIG. \ref{fig:DOS} and details of band structure in FIG. \ref{fig:bands} respectively. The DOS has a significant value at the E$_F$ signifying a metallic state. The s-band of N atom lies deep inside at around -17 eV, with a width of 1.7 eV, having symmetric up and down spin bands.

The Fe d-band shows a distinct 3-fold and 2-fold character. The broad 3-fold Fe bands dominate the E$_F$ and have nearly only Fe-d character. The majority and minority bands have bandwidths 7.25 and 7.53 eV respectively, hence they are not strictly rigid band shifts. An exchange splitting of 0.79 eV is obtained from the difference of the band centers of the majority and minority bands. The 3-fold Fe-d bands at the E$_F$ contribute to Fe-Fe bonding, as well as the magnetic moment, since the majority and minority bands are filled to different extents. Thus, the magnetic moment on the Fe site is 1.528 $\mu_\text{B}$.

The 2-fold Fe d-bands are narrower (bandwidth of 3.184/3.185 eV for up/down spins) and located at around -7 eV and show hybridization with N-p band. These 2-fold Fe d-bands are responsible for Fe-N bonding. The results of density of states are very clearly depicted in band structures shown in FIG. \ref{fig:bands}. The band structure shows filled N-p band around -6 to -8 eV while Fermi level characteristics are solely due to Fe d-band. 

\begin{figure*}
\includegraphics[width=17.2cm]{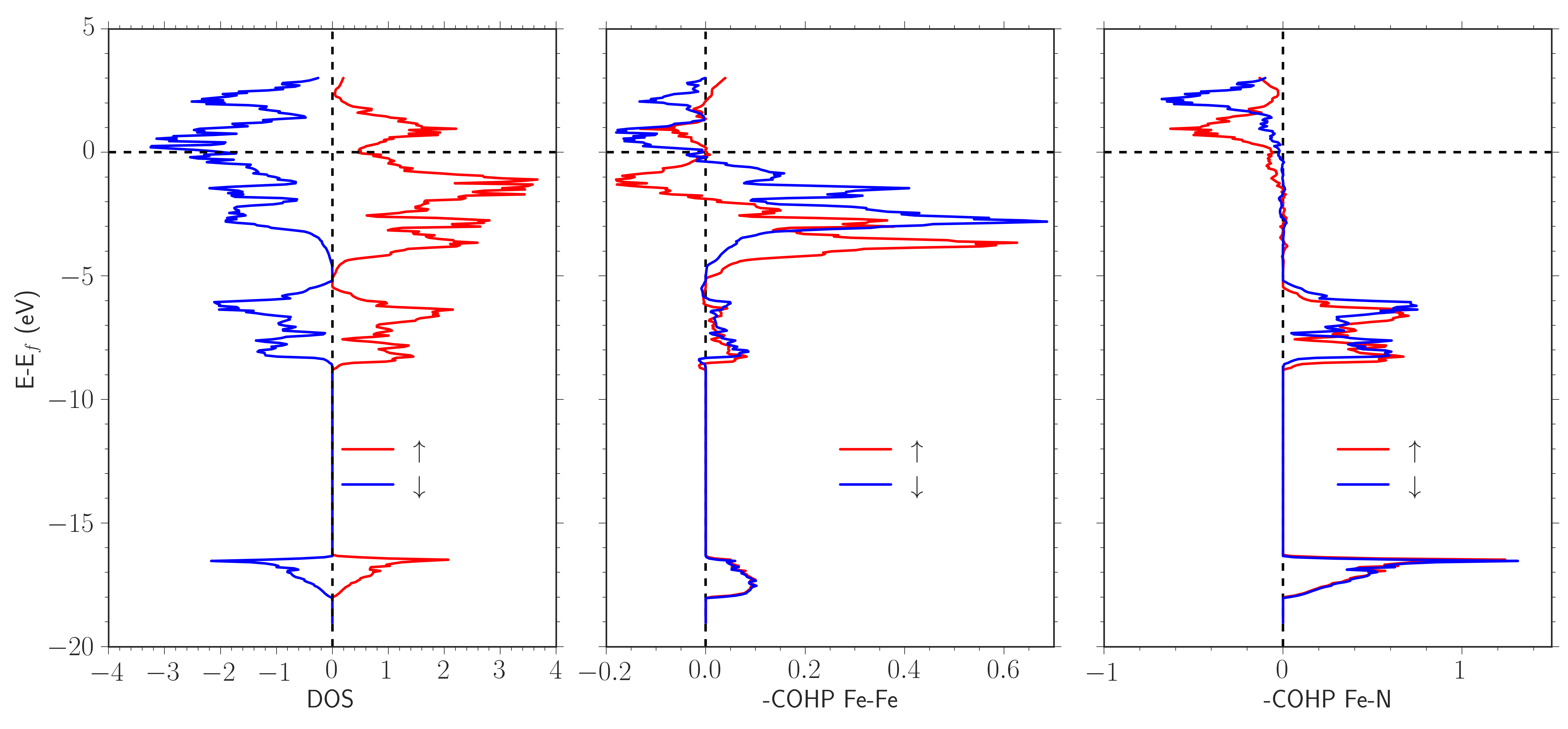}
\caption{(Color online) DOS and COHP curves for closest Fe-Fe and Fe-N atom pairs}
\label{fig:COHP}
\end{figure*}

Crystal Orbital Hamilton Population analysis partitions the band structure energies into contributions coming from bonding, anti-bonding or non-bonding interactions. Shown here in FIG. \ref{fig:COHP}, the DOS curve with –COHP curves for the closest Fe-N and Fe-Fe atom pairs. A positive value signifies a bonding contribution while a negative value is for an anti-bonding contribution. We see that COHP curve supports our claims that the 2-fold band near the E$_F$ is Fe-N bonding overlap and that the 3-fold band at the E$_F$ is primarily Fe-Fe bonding interaction.

\begin{figure*}
\includegraphics[width=17.2cm]{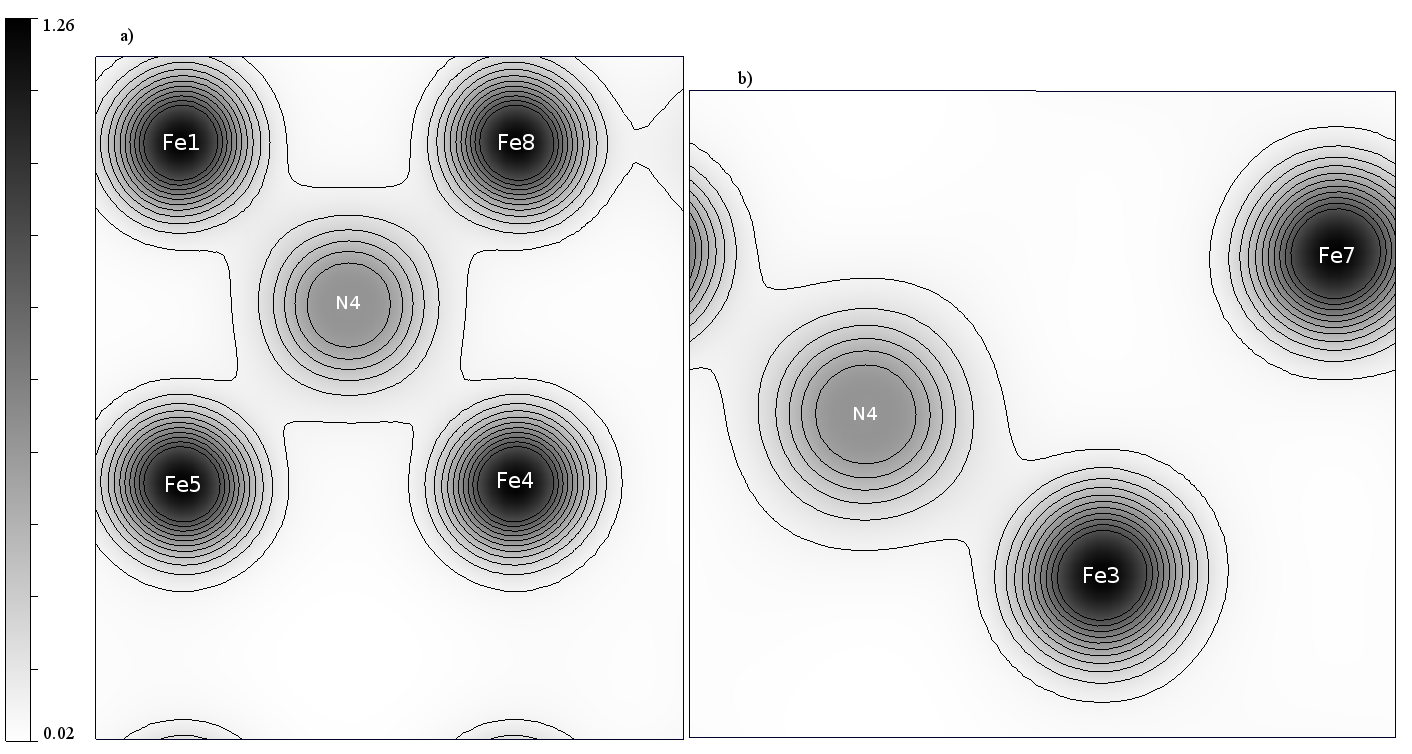}
\caption{Charge Density Contour a) along and b) perpendicular to the octahedral plane of the 'N4' octahedra depicted in FIG \ref{fig:unitcell}. Contour lines equally spaced in steps of 0.08 e/\AA$^3$}
\label{fig:contour}
\end{figure*}

The charge density contours along and perpendicular to the base of octahedra containing N4 are shown in FIG \ref{fig:contour} (a) \& (b). We find from the charge density contours that Fe atoms interact only via N atoms, i.e. there is no direct Fe-Fe charge density overlap. Also note that the overlap seen in the upper-right corner FIG. \ref{fig:contour}a, corresponds to the overlap of Fe8 with N1 atom and not with the period image of Fe1.

Thus, the contours depict two types of Fe-Fe interactions: 
\begin{itemize}
\item In the plane of the octahedra with a Fe-N-Fe bond angle of $\sim$ 90$^o$, i.e. the $90 ^o$ superexchange interaction
\item Perpendicular to the octahedral plane with a Fe-N-Fe bond angle of $\sim$ 180$^o$, i.e. the $180 ^o$ superexchange interaction
\end{itemize}
As shall be seen later, the Fe-N-Fe with bond angles $\sim$ 90$^o$ interact ferromagnetically, while Fe-N-Fe with bond angles  $\sim$ 180$^o$ have AFM character. 

\section{Magnetic State}
The results on the magnetic state of Fe$_2$N have been contrasting and contradicting. Although earliest experimental measurements by Bridelle\cite{bridelle} report the material to be FM, a few experimental reports\cite{bainbridge_mossbauer_1973, hinomura_study_1998, kano_magnetism_2001} have been contradictory. Bainbridge et al do not observe any magnetic ordering down to 5K and report PM while Hinomura et al and Kano et al find $\zeta$-Fe$_2$N to be AFM with a Neel temperature of ~9-11K.

The results of Hinomura \footnote{see reference \onlinecite{hinomura_study_1998}, FIG. 1} are based on $^{57}$Fe M{\"o}ssbauer measurements of $\zeta$-Fe$_2$N at 4.2K in presence and absence of 7T applied magnetic field. The M{\"o}ssbauer measurements can never really conclusively differentiate between AFM state and FM state,\cite{bhargava_phase_2004, bhargava_phase_2005, bhargava_critical_2007, patange_random_2015} unless supported with measurements on temperature dependence of susceptibility, and hence the results of Hinomura could be well be of the FM state.

Similarly the works of Kano\cite{kano_magnetism_2001} et. al. also inconclusively prove AFM ordering with Neel temperature near 11K in case of Fe$_{1.99}$N and Fe$_{2.01}$N. The Fe-N phase diagram shows that $\zeta$-Fe$_2$N is strictly a line phase, while non-stoichiometric deviations would form the $\epsilon$ phase which is a crystallographically distinct from $\zeta$ phase.\cite{lehrer_2002} It is possible that the AFM state of Fe$_2$N observed by Hinomura et. al. was the non-stoichiometric $\epsilon$ phase.

Hence, the experimental observation of AFM state of $\zeta$-Fe$_2$N is inconclusive. There are no experimental measurements which conclude affirmatively that the state is PM. Arrott plots and inverse susceptibility plots have to be investigated with care for transition temperature as well as the nature of the coupling. Experimental measurements of susceptibility as a function of temperature could yield misleading results because actual plots of inverse susceptibility with temperature are never perfectly linear, \cite{curie_weiss_plot} particularly for itinerant ferromagnets. Hence, estimation of y-intercept leading to either ferro, para or antiferro magnetic ordering could be incorrect particularly in cases where the magnetic transition temperature is low.

As regards computational first principle studies, a few reports\cite{matar_electronic_1993,fang_role_2011} conclude the magnetic state to be FM while a couple of reports \cite{eck_theoretical_1999, chen_phase_2015} claim PM ground state. It is intriguing that none of the theoretical studies conclude the ground state to be AFM and still further none of them evaluate Stoner criterion to support their claims for a FM ground state.

\subsection*{Stoner estimates:}
The characteristic properties of magnetic state, such as Curie temperature (T$_C$), computed within the framework of band theory are grossly overestimated\cite{hubbard_magnetism_1979}. Within the band model for magnetism in the spin density functional formalism, the net magnetization per atom ($\Delta m$) is given by\cite{gunnarsson_band_1976} 

\begin{equation}
\Delta m = \int_{WS\ cell} \left[ \rho_+(\textbf{r}) - \rho_-(\textbf{r}) \right]/ d^3\textbf{r} \equiv \int_{WS\ cell} m(\textbf{r}) d^3\textbf{r}
\end{equation}

We follow here notations given in the paper by O. Gunnarsson\cite{gunnarsson_band_1976}, where $\rho$ represents the spin density, +/- the spin index, the quantity m(r) represents the fractional spin polarization  and the integration is over the Wigner Seitz cell.

Within first order perturbation arguments, $\Delta m$ is related to the band splitting as,
\begin{equation}
\Delta \epsilon_{kn} = \epsilon_{kn+} - \epsilon_{kn-} = \Bra{kn} v^{xc}_{+} - v^{xc}_{-} \Ket{kn} = \Delta m \Bra{kn} v^{xc}_{(1)} \Ket{kn} \equiv - I (\epsilon_{kn})\Delta m
\end{equation}

where $\epsilon_{kn}$ are energy levels characterized by orbitals $\Ket{kn}$, $v^{xc}$ is the exchange correlation potential within the local spin density approximation (LSDA) and I is the stoner exchange parameter. The generalized Stoner parameter I($\epsilon$) \cite{gunnarsson_band_1976} is given by the expression

\begin{equation}\label{eq:genStoner}
I(\epsilon) = - \int_{0}^{r_{WS}} r^2 dr \left( \frac{\mu^k(r_s)}{6\pi} \right) \delta(r_s) \frac{\phi^2_2(r, \epsilon) \phi^2_2(r, \epsilon_F)}{\rho(r)}
\end{equation}

I($\epsilon$), which is characteristic for a material, in the above equation depends implicitly on k through energy. It is physically argued that the states under consideration are of d-type only. This is a good approximation since the non-spherical part of spin polarization give only a small contribution to the band splitting. The equation for Curie temperature is given by,

\begin{equation}\label{eq:stonerTc}
I(E_f) \times \int_{-\infty}^{\infty} \frac{\partial f(E)}{\partial E} \rho(E)dE + 1 = 0
\end{equation}

Where f(E) is the well-known Fermi-Dirac distribution function, $f(E)=(1+ e^{(E-E_f)/kT})^{-1}$, I, $\rho$(E), $E_f$, k and T are the stoner parameter, density of states per atom, Fermi energy, Boltzmann constant and absolute temperature respectively.
This reduces to Equation \ref{eq:stonerCritera} for T=0, which is nothing but the Stoner criterion for ferromagnetism.

\begin{equation}\label{eq:stonerCritera}
I(E_f) \times \rho(E_f) > 1
\end{equation}

The exchange splitting ($\Delta$) in the stoner model can be expressed as:
\begin{equation}\label{eq:stonerI}
\Delta = I \times M
\end{equation}

where I is the stoner parameter and M is the local magnetic moment. We obtain $\Delta$ from the difference of  band center position of 3-fold for majority and minority d-bands. It is essential that band-centers and not peak positions are considered when computing $\Delta$, as the bands here are not strictly symmetrical and have varying band widths\cite{kubler_magnetic_1981}. Consideration of peak positions leads to a much higher value of $\Delta$ and correspondingly higher I. The value of Stoner parameter thus estimated from Equation \ref{eq:stonerI} is 0.523.

The stoner criterion and $T_C$ computed using equation \ref{eq:stonerTc} are tabulated in Table \ref{table:stoner}.The product was found to be 1.404, thus $\zeta$-Fe$_2$N satisfies the stoner criterion for ferromagnetism, though not by a very large margin (when compared to 3.277 obtained for bcc-Fe). This is in agreement with our energetics calculations which place the FM state to be marginally more stable than the AFM or NM states. This result should also explain why the experimental studies so far have come up with conclusions of PM/AFM states.

\begin{figure}
\includegraphics[width=9cm]{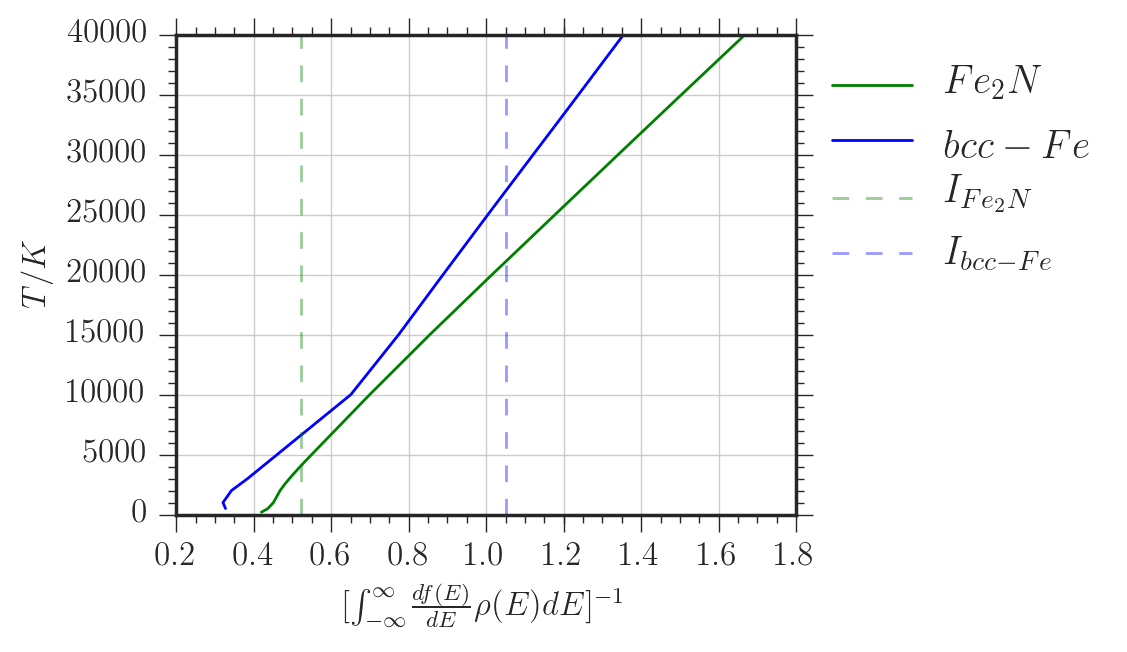}
\caption{(Color online) Estimation of Curie Temperature}
\label{fig:Curie}
\end{figure}

The $T_C$ values, under the band theory model tabulated in Table \ref{table:stoner}, are obtained graphically as per Equation \ref{eq:stonerTc}, from a plot of T vs $ \left[ \int_{-\infty}^{\infty} \frac{\partial f(E)}{\partial E} \rho(E)dE \right]^{-1} $ depicted in FIG \ref{fig:Curie}.

\begin{table}
\caption{Stoner parameter and Curie temperature predicted by Stoner model}
\label{table:stoner}
\begin{tabular}{|c|c|c|c|}
\hline 
System & I(eV) & $I\rho(E_{f})$ & $T_{C}$ (K)\tabularnewline
\hline 
\hline 
$\zeta-Fe_{2}N$ & 0.5229 & 1.404 & 4100\tabularnewline
\hline 
BCC-Fe & 1.0504 & 3.2772 & 27000\tabularnewline
\hline 
\end{tabular}
\end{table}

\subsection*{Estimation of transition temperature and coupling constant}

In the Stoner model, the temperature dependence of magnetic properties is attributed solely to the temperature dependence of Fermi distribution, thereby resulting in exaggerated magnetic properties of materials. This is due to the fact that it includes only spin flip transitions, i.e. from up-spin band to down spin band as temperature increases. In contrast to the Stoner's band model for magnetism, Heisenberg's model presents an atomistic picture for magnetic interaction and concomitant transition temperature. At zero temperature and for zero applied external magnetic field, the difference between spin polarized total electronic energy (H$^{sp}$) and non-spin polarized total electronic energy (H$^{nsp}$) are approximately equal to magnetic energy given by the Heisenberg model. That is,
\begin{equation}
H^{sp} - H^{nsp} = -\sum_i \sum_j \lambda_{ij} \textbf{J}_i . \textbf{J}_j
\end{equation}
where $\bf J_{i}$ is the total angular momentum on the $i^{th}$ site and $\lambda_{ij}$ is exchange coupling constant between the sites. Since the structure has translational periodicity the sum over "i"s in the above equation correspond to sum over all Fe sites in the unit cell and the energy difference $H^{sp} - H^{nsp}$ is estimated in eV/unit cell. Furthermore, in the present discussion we assume only near neighbor interaction between Fe sites and we ignore corrections due to magnetic anisotropy and assume the easy axis of magnetization to be along the c-axis. In the previous section we observed that that there are two types of Fe-Fe interactions, which maybe characterized by two coupling constants (say $\lambda_{90}$ and $\lambda_{180}$). To determine the values of these coupling constants we construct different spin configurations.

\subsubsection{Alternating layers along c-axis}
\begin{figure*}
\includegraphics[width=17.2cm]{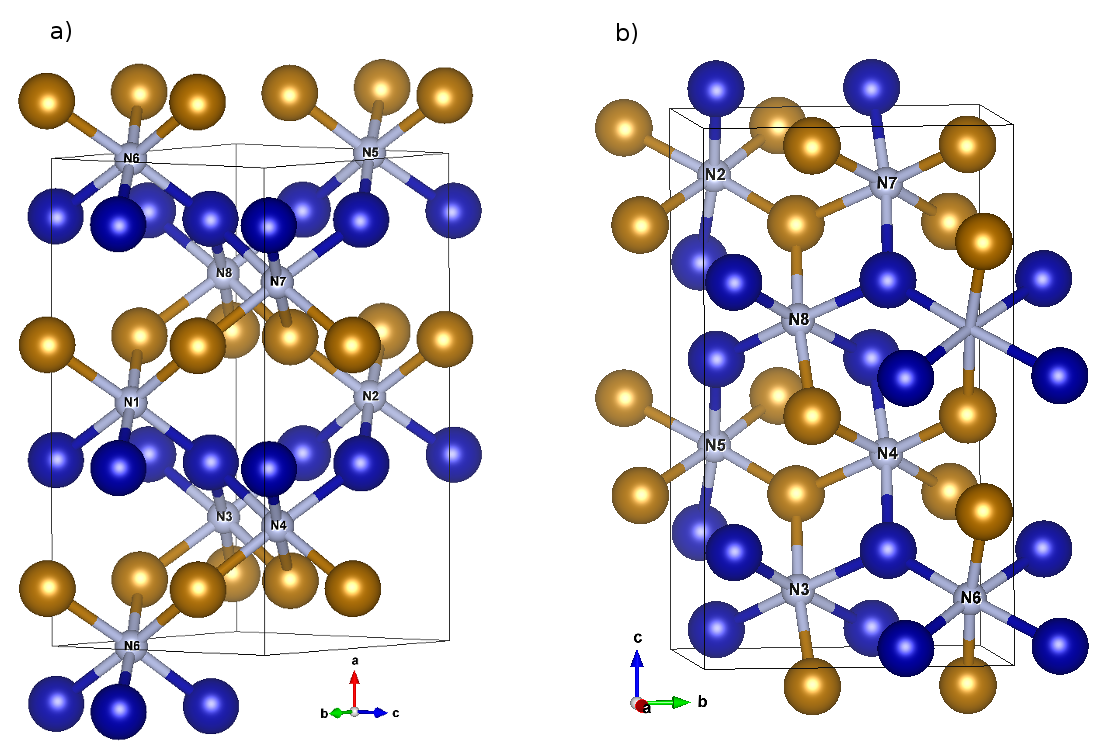}
\caption{(Color online) Two spin configurations with with gold and blue balls representing Fe atoms with opposite spin orientation. a) Cell doubled along a-axis b) Cell doubled along c-axis.}
\label{fig:spinconfigs}
\end{figure*}

Consider first the case where we double the cell along c-axis and adjacent layers of Fe have opposite spin orientation (refer FIG \ref{fig:spinconfigs}b). We see in this configuration that for any octahedra, opposite vertices have balls of similar color, i.e. all $180 ^o$ bonds are between Fe pairs having parallel orientation. Also note, starting at any one blue (or gold) ball that it has four balls of same color with bond angle ($90 ^o$) and eight balls of different color with bond angle ($90 ^o$).

To summarize, out of the 15 nearest neighbor (Fe-Fe) pairs, we have:
\begin{itemize}
\item Three $180 ^o$ Fe-Fe pairs with a parallel magnetic orientation
\item Four $90 ^o$ Fe-Fe pairs with parallel magnetic orientation
\item Eight $90 ^o$ Fe-Fe pairs with anti-parallel magnetic orientation
\end{itemize}

Since we have equal number of spin-up and spin-down Fe atoms here, this is an AFM configuration whose spin Hamiltonian ($H_{c}^{AFM}$) may be written as:

\begin{equation}
H^{AFM}_{c} = -\sum_i \sum_j \lambda_{ij} \textbf{J}_i . \textbf{J}_j
\end{equation}

At zero Kelvin, the spins are either going to be parallel or anti-parallel, hence we consider only the magnitude of $\textbf{J}$. Now considering only nearest neighbor interactions and all Fe atoms to be identical, we obtain for the unit cell containing 16 Fe atoms:
\begin{equation}
H^{AFM}_{c} = - 8 \sum_j^{15} \lambda_{1j} |\textbf{J}_1| |\textbf{J}_j|
\end{equation}
The sum now runs over the 15 nearest neighbor interactions, this can further be expanded as:
\begin{multline}
H^{AFM}_{c} = - 8 \left[ 3 \lambda_{180} |\textbf{J}||\textbf{J}| + 4 \lambda_{90} |\textbf{J}||\textbf{J}| + 8 \lambda_{90} |\textbf{J}|(-|\textbf{J}|) \right] \\= -24 \lambda_{180} |\textbf{J}|^2 + 32 \lambda_{90} |\textbf{J}|^2
\end{multline}

Now consider the same cell with \emph{all} spins oriented in parallel, we have now, the Hamiltonian for the FM configuration ($H_{c}^{FM}$):
\begin{multline}
H^{FM}_{c} = - 8 \left[ 3 \lambda_{180} |\textbf{J}||\textbf{J}| + 12 \lambda_{90} |\textbf{J}||\textbf{J}| \right] \\= -24 \lambda_{180} |\textbf{J}|^2 - 96 \lambda_{90} |\textbf{J}|^2
\end{multline}
Taking a difference, we obtain:
\begin{equation}\label{eq:lambda90}
H^{FM}_{c} - H^{AFM}_{c} = -128 \lambda_{90} |\textbf{J}|^2
\end{equation}

We obtain $|\textbf{J}|$ as, $|\textbf{J}| = \frac{-1}{g_J} |\bm{\mu}|$, where $g_J$ is the Land{\'e} g-factor (here, 1.5 for Fe), $\mu$ is the magnetic moment obtained from VASP calculation.

Substituting the energies of the FM and AFM spin configurations computed from VASP into Eq \ref{eq:lambda90} we obtain,

\begin{equation}
\lambda_{90} = 0.00587\ eV = 5.96 \ meV
\end{equation}

\subsubsection{Alternating layers along a-axis}
In this configuration (refer FIG \ref{fig:spinconfigs}a), for any octahedra, all opposite vertices have balls of different color, i.e. all $180 ^o$ bonds are between Fe pairs having anti-parallel orientation. Also note, starting at any one blue (or gold) ball that it has six balls each of the blue and gold color with bond angle ($90 ^o$).

To summarize in this configuration, out of the 15 nearest neighbor (Fe-Fe) pairs, we have:
\begin{itemize}
\item Three $180 ^o$ Fe-Fe pairs with anti-parallel magnetic orientation
\item Six $90 ^o$ Fe-Fe pairs with parallel magnetic orientation
\item Six $90 ^o$ Fe-Fe pairs with anti-parallel magnetic orientation
\end{itemize}

The spin Hamiltonian ($H^{AFM}_{a}$) may be written as discussed in the previous section as:
\begin{multline}
H^{AFM}_{a} = - 8 [\ 3 \lambda_{180} |\textbf{J}|.(-|\textbf{J}|) + 6 \lambda_{90} |\textbf{J}||\textbf{J}| \\+ 6 \lambda_{90} |\textbf{J}|.(-|\textbf{J}|) ] = 24 \lambda_{180} |\textbf{J}|^2
\end{multline}

Now consider the same cell with \emph{all} spins oriented in parallel, we have now the spin Hamiltonian ($H^{FM}_{a}$):
\begin{multline}
H^{FM}_{a} = - 8 \left[ 3 \lambda_{180} |\textbf{J}|.(|\textbf{J}|) + 6 \lambda_{90} |\textbf{J}|.|\textbf{J}| + 6 \lambda_{90} |\textbf{J}||\textbf{J}| \right] \\= - 24 \lambda_{180} |\textbf{J}|^2 - 96 \lambda_{90} |\textbf{J}|^2
\end{multline}

Taking a difference of the two,
\begin{equation}
H^{FM}_{a} - H^{AFM}_{a} = -48 \lambda_{180} |\textbf{J}|^2 -96 \lambda_{90} |\textbf{J}|^2
\end{equation}

Substituting the computed energy difference from VASP and value of $\lambda_{90}$ computed in the previous section, we obtain
\begin{equation}
\lambda_{180} = -0.01040\ eV = -10.57 \ meV
\end{equation}
We see that $\lambda_{90}$ is positive and thus the coupling is FM as opposed to the negative sign of $\lambda_{180}$ which points to an AFM coupling, which are in accordance with Goodenough-Kanamori-Anderson (GKA) rules\cite{kanamori_superexchange_1959}. Also the magnitude of the AFM coupling constant is almost twice that of the FM coupling constant. However, since each Fe site is ferromagnetically coupled to twelve nearest neighbor Fe sites and also simultaneously coupled anti-ferromagnetically with three nearest neighbor Fe atoms with a Fe-N-Fe bond angle of 90 and 180 degrees respectively, the net coupling constant on any Fe atom can be expressed as a weighted average, i.e.

\begin{equation}
<\lambda> = \frac{3 \lambda_{180} + 12 \lambda_{90}}{15} = 0.00265\ eV = 2.65 \ meV
\end{equation}

Inspite of the larger magnitude of AFM coupling constant, the overall average coupling constant turns out to be positive because of the greater number of FM interactions. Characterizing properties of magnetic state, such as Curie constant (C) $\&$ Curie temperature (T$_C$), are given by
\begin{equation}
C = \frac{n \mu_0 \mu^2}{3k_B}
\end{equation}
where, n is the number of Fe atoms per unit volume, $\mu_0$ is the permeability of vacuum, $\mu$ is the magnetic moment of Fe and $k_B$ is the Boltzmann constant, z is the number of nearest neighbors. 

From mean field theory we have \cite{kittel},
\begin{equation}
T_c = \frac{2z J(J+1) \ \lambda}{3k_B} = \frac{2z}{3k_B} \frac{\mu^2}{g_J^2} \lambda
\end{equation}
where, $T_C$ is the Curie temperature, z the number of nearest neighbors, $\mu$ the magnetic moment in bohr magneton and $g_J$ is the Land{\' e} g-factor. For d-band magnetism as in $Fe_2N$, using the value of $g_J$=1.5 for Fe, we have a value of C = 0.4323 K \& T$_C$ = 479 K. 

Table \ref{table:summary} summarizes the results obtained in the present work and compares them with those published earlier. The magnetic moment on the Fe site as obtained from first principles spin polarized calculations match well with those published elsewhere\cite{fang_role_2011}. Furthermore, Fang\cite{fang_role_2011} et. al \& other theoretical investigators have neither computed Stoner parameter nor determined coupling constant and transition temperature. Our computed Curie temperature, within Heisenberg model, is the first theoretical estimate for Fe$_2$N and although it overestimates the observed experimental results, it is significantly smaller than the value computed within Stoner band model. However, we must point out that our estimates of Curie temperature are based only on near neighbor interactions between Fe sites and the contribution from next nearest neighbor and beyond may be important for more accurate estimation of T$_c$, in light of the fact that the compound is metallic. 
    
\section{Conclusions}
In this paper we discuss the magnetic ground state of $\zeta$-Fe$_2$N against the backdrop of a long list of conflicting computational and experimental results. We find the experimental studies which observe an AFM ground state to be inconclusive due to the limitations in M{\"o}ssbauer technique and also absence of data on temperature dependence of magnetization. Our computational study suggests a FM ground state with a magnetic moment on Fe site consistent with previous computational studies. An evaluation of Stoner criterion further supports our claim for a FM ground state, although it yields a grossly exaggerated value for T$_C$. Furthermore, we observe that the nature of the Fe-N-Fe bonding is such that there exists a superexchange coupling of both AFM and FM type between the Fe-Fe pairs mediated by nitrogen atoms. We thus have, around every Fe atom, twelve FM Fe-N-Fe pairs and three AFM Fe-N-Fe pairs, resulting in a net FM ordering in $\zeta -$Fe$_2$N. For the first time we have estimated exchange coupling constant, Curie temperature \& Stoner parameter for this phase. Although our estimate for Curie temperature (T$_c$ = 479 K, Heisenberg model) is higher than experimentally observed values,  inclusion of next nearest neighbor interactions and beyond could possibly result in a more realistic estimate.

\begin{acknowledgments}
KSR thanks Shri Prathap Reddy K, Dr. K. C. Rao, Dr. C. L. Prajapat, Dr. T. V. C. Rao and Dr. S. C. Gadkari for their support and encouragement.
\end{acknowledgments}
    
\begin{table*}
\caption{Summary of Results}
\begin{tabular}{|c|c|c|c|c|c|c|}
\hline 
\multirow{2}{*}{Work} & Mag. Moment  & Exchange & Stoner  & Stoner & Critical & Magnetic State \tabularnewline
 & on Fe ($\mu_B$) &  Splitting (eV) & Parameter(I) &  Criterion & Temperature  (K) & Concluded\tabularnewline
\hline 
\hline 
Present Work\footnote{\label{compwork}Computational Work} & 1.528 & 0.799 & 0.5229 & 1.404 & T$_C$ = 4100\footnote{\label{stoner}Stoner Model}, 479\footnote{\label{heisenberg}Heisenberg model} & FM\tabularnewline
\hline 
Chen (2015)\footref{compwork} & 1.38 & - & - & - & - & PM\tabularnewline
\hline 
Fang (2011)\footref{compwork} & 1.5 & $\sim$2.0 & - & - & - & FM\tabularnewline
\hline 
Kano(2001)\footnote{\label{exptwork}Experimental Work} & - & - & - & - & $T_{N}$=11-12K & AFM\tabularnewline
\hline 
Sifkovits (1999)\footref{compwork} & 1.43 & 1.2 & - & - & - & No Comment\tabularnewline
\hline 
Eck (1999)\footref{compwork} & 1.5 & - Not reported & - & - & - & PM\tabularnewline
\hline 
Hinomura (1998)\footref{exptwork} & - & - & - & - & $T_{N}$=9K & AFM\tabularnewline
\hline 
Matar (1993)\footref{compwork} & 1.47 & Not reported & - & - & - & FM\tabularnewline
\hline 
Bainbridge (1973)\footref{exptwork} & - & - & - & - & - & PM\tabularnewline
\hline
Bridelle (1955)\footref{exptwork} & - & - & - & - & - & FM\tabularnewline
\hline  
\end{tabular}
\label{table:summary}
\end{table*}


\bibliography{fe2nbib_noupload}

\end{document}